\documentclass[comsoc, journal]{IEEEtran}

\usepackage{graphicx}
\usepackage{cite}
\usepackage{filecontents}
\usepackage[square,sort,comma,numbers]{natbib}
\usepackage{tabularx,booktabs}
\usepackage{multirow}
\usepackage{makecell}
\setcellgapes{3pt}

\usepackage{lipsum}
\usepackage{color, xcolor}
\usepackage{algpseudocode,algorithm,algorithmicx,float}

\def \revision{\color{black}}

\begin{document}

\title{FoggyEdge: An Information Centric Computation Offloading and Management Framework for Edge-based Vehicular Fog Computing}

\author{Muhammad Atif Ur Rehman,
        Muhammad Salahuddin,~\IEEEmembership{Student Member,~IEEE,}
        Spyridon Mastorakis,~\IEEEmembership{Member,~IEEE,}
        and~Byung-Seo Kim,~\IEEEmembership{Senior Member,~IEEE,}
\thanks{Muhammad Atif Ur Rehman is with the Department of Computing and Mathematics, Manchester Metropolitan University, UK, Muhammad Salahuddin is with the Electronics and Computer Engineering Department, Hongik University, South Korea, Spyridon Mastorakis is with the College of Engineering, University of Notre Dame, USA. Byung-Seo Kim is with Softwae and Communication Engineering Department, Hongik University, South Korea.}

\thanks{Manuscript received April 19, 2015; revised August 26, 2015.}}

\markboth{Journal of \LaTeX\ Class Files,~Vol.~14, No.~8, August~2015}%
{Shell \MakeLowercase{\textit{et al.}}: Bare Demo of IEEEtran.cls for IEEE Journals}

\makeatletter
\def\ps@IEEEtitlepagestyle{%
  \def\@oddfoot{\mycopyrightnotice}%
  \def\@oddhead{\hbox{}\@IEEEheaderstyle\leftmark\hfil\thepage}\relax
  \def\@evenhead{\@IEEEheaderstyle\thepage\hfil\leftmark\hbox{}}\relax
  \def\@evenfoot{}%
}
\def\mycopyrightnotice{%
  \begin{minipage}{\textwidth}
  \centering \scriptsize
978-1-5386-5541-2/18/\$31.00 ©2018 IEEE. This paper has been accepted for publication by the IEEE Intelligent Transportation System Magazine. The copyright is with IEEE and the final version will be published by IEEE.
  
  \end{minipage}
}
\makeatother
\maketitle

\begin{abstract}
The recent advances aiming to enable in-network service provisioning are empowering a plethora of smart infrastructure developments, including smart cities, and intelligent transportation systems. Although edge computing in conjunction with roadside units appears as a promising technology for proximate service computations, the rising demands for ubiquitous computing and ultra-low latency requirements from consumer vehicles are challenging the adoption of intelligent transportation systems. Vehicular fog computing which extends the fog computing paradigm in vehicular networks by utilizing either parked or moving vehicles for computations has the potential to further reduce the computation offloading transmission costs. Therefore, with a precise objective of reducing latency and delivering proximate service computations, we integrated vehicular fog computing with roadside edge computing and proposed a four-layer framework named FoggyEdge. The FoggyEdge framework is built at the top of named data networking and employs microservices to perform in-network computations and offloading. A real-world SUMO-based preliminary performance comparison validates FoggyEdge effectiveness. {\revision Finally, a few future research directions on incentive mechanisms, security and privacy, optimal vehicular fog location, and load-balancing are summarized.}
\end{abstract}


\IEEEpeerreviewmaketitle

\section{Introduction}

\IEEEPARstart{T}The autonomous and/or smart computerized vehicles constitute a considerable portion of the Intelligent Transportation System (ITS), enabling ubiquitous computing and pervasive connectivity through various onboard computational units and wireless communication technologies including ultra-wideband Bluetooth (UWB), Wi-Fi, dedicated short-range communication (DSRC), and cellular vehicle to everything (C-V2X), to name a few, subsequently paving the way for the complete realization of the Internet of ITS (IoITS) \cite{autonomousvehicleICN}. In the vision of IoITS, smart vehicles will fetch and process complex multimodal visual content in real-time for better contextual awareness, consequently, avoiding fatal crashes, improving the comfort of passengers, and the safety of people on the sidewalks, and vehicles on the roads. These complex computational processing may require substantial computation power and stringent ultra-low latency, thus posing significant challenges to today’s contemporary vehicles that solitarily are unable to meet required computational demands \cite{mecvehicle}. 

To meet the ever-increasing computation requirements of consumer vehicles, remote cloud computation is initially adopted in vehicular networks, allowing the consumer vehicles to offload the computations—if the local resources are exhausted, on a cloud node located far away from the requesting vehicle. Remote cloud offloading, undeniably, relieved the computation burden on consumer vehicles, and improved the computational resource utilization, nonetheless, dealing with the strict latency requirement is a daunting challenge. To overcome the shortcomings of remote cloud offloading, edge computing is introduced as a promising solution in the recent past, extending the features of cloud nodes, such as computational resources, storage capacity, and services code at vehicular terminals, empowering smart vehicles to offload the tasks in proximity.

Although edge-integrated vehicular terminals, often termed vehicular edge, are to some extent resolving both computations and latency requirements, the tremendous increase in smart vehicles specifically in urban areas is putting a significant burden on the vehicular edge, consequently forcing edge nodes to offload tasks to remote clouds through the Internet. Since cloud nodes are usually located far away from the vehicular edge, yet again smart vehicles experience delays. Moreover, in existing vehicular edge-to-cloud computations offloading, the content transmission follows the {\revision inefficient } address-based TCP/IP protocol, which further increases the delays as a result of congestion in dense traffic scenarios \cite{connectedvehicles}. Therefore, the existing computation offloading schemes on top of traditional networking architectures, which often exploit the synergy of edge and cloud infrastructure, are now confronting substantial challenges in handling the rising concerns.   

To efficiently address these issues, this article envisions the idea of FoggyEdge, a vehicular edge fogged by parked vehicles to perform computations. The FoggyEdge framework employs a fog of resource-rich vehicles—in the proximity of vehicular edge, which is expected to have the high computational power and large storage capacity. By introducing vehicular fog, we designed a four-layer computation offloading framework consisting of 1) the things layer, 2) the vehicular edge layer, 3) the vehicular fog layer and 4) the cloud layer. Our unique framework design follows the named data networking (NDN) principles as an underlying communication architecture. Moreover, to migrate computational service code among various layers, we use microservices, decomposing the bulky monolithic services into small, manageable, and independent chunks of service code. Besides, to validate the computation offloading request, we proposed a microservice access management mechanism that enables the vehicles to consume the microservices only if they have legitimate rights. Furthermore, the proposed framework proposes a vehicular fog management mechanism that efficiently manages the resources of vehicles inside the vehicular fog. Finally,  a preliminary performance comparison demonstrates the effectiveness of the proposed framework in terms of computation satisfaction delay.

In summary, the following are the key contributions of this paper.
\begin{itemize}
\item	The FoggyEdge features a unique four-layered vehicular edge and fog computations offloading framework, exploiting NDN as underlying communication architecture and aiming at enabling proximate computation to reduce the latency that is incurred otherwise in traditional approaches. 
\item	FoggyEdge employs microservices architecture to migrate computational services code among various layers, consequently reducing the bandwidth consumption and the chances of network congestion, and delays. 
\item	The FoggyEdge introduces a microservice access management mechanism to validate the computation offloading request, enabling the microservices to be consumed only if the requesting vehicle has legitimate rights. 
\item	The FoggyEdge proposes vehicle orchestration and admission process by maintaining the vehicular fog resource access table (VF-RAT) at the vehicular fog gateway (VFG) to efficiently manage the computational resources and in-progress computations running inside the vehicular fog. 
\end{itemize}

{\revision The proposed integration of vehicular fog with an edge by considering microservices for computations and service migration, and the NDN as an underlying communication architecture has indeed all the potential to improve the autonomous and modern vehicles transportation system. Nevertheless,  further research efforts with a focus on developing efficient incentive mechanisms, security and privacy mechanism, optimal vehicular fog location, and load-balancing, is required to further improve the FoggyEdge system. These aforementioned open future research directions are highlighted at the end of this paper.}

The rest of the paper is organized as follows. Section \ref{sect-background} sheds light on fundamental building blocks. Related studies are presented in Section \ref{sect-related}. Comprehensive details on the proposed FoggyEdge framework are outlined in Section \ref{sec-proposed}. The preliminary comparative performance evaluation study is presented in Section \ref{sec-performance}. Section \ref{sec-future} highlights the potential research challenges and possible future directions. Finally, Section \ref{sec-conclusion} concludes the paper.

\section{A PRIMER OF FUNDAMENTAL BUILDING BLOCKS }
\label{sect-background}
The key building blocks on which the FoggyEdge framework is built are 1) vehicular fog computing: aiming to provide proximate computations, 2) microservices: for efficient computation and services migrations, and 3) NDN: as underlying communication architecture. The following subsections shed light on the key aspects of these building blocks.

\begin{table*}[!t]
\label{tabcompstudies}
	\centering
	\caption{Summary of Related Studies}
	\begin{tabular}{|c|c|c|c|c|c|}
		\hline
		\textbf{References}                       & \multicolumn{1}{m{7cm}|}{\textbf{Objective}}     & \multicolumn{1}{m{2cm}|}{\textbf{Named-Centric Communication}}& \multicolumn{1}{m{2cm}|}{\textbf{Microservice Architecture}} & \multicolumn{1}{m{1.5cm}|}{\textbf{Access Mechanism}}& \multicolumn{1}{m{1.5cm}|}{\textbf{VFC \& EC Integration}}                                                                \\ \hline
		
		\cite{mecvehicle}                       & \multicolumn{1}{m{7cm}|}{Cloud-based mobile edge computing offloading framework for vehicular networks}      & {x} & {x} & {x} & {x}                                                                                                        \\ \hline
		
		\cite{MEC}                       & \multicolumn{1}{m{7cm}|}{Four layered multi-access edge computing and fog computing architecture }      & {x} & {x} & {x}& \checkmark                                                               \\ \hline
		
		\cite{VFCTraffic}    & \multicolumn{1}{m{7cm}|}{Three-layered VFC framework for latency reduction \& computations offloading }      & {x} & {x} & {x} & \checkmark                                                                                                         \\ \hline
		
		\cite{CVC} & \multicolumn{1}{m{7cm}|}{Task scheduling mechanism for computation efficiency improvement in mobile vehicular clouds}      & {x} & {x} & {x}& \checkmark                                                                                                         \\ \hline
				\cite{SDNEV} & \multicolumn{1}{m{7cm}|}{SDN based mobile edge computing and fixed edge computing integration framework}      & {x} & {x} & {x} & \checkmark                                                                                                         \\ \hline

		FoggyEdge                       & \multicolumn{1}{m{7cm}|}{Four layered named-centric vehicular fog and edge computing framework for latency reduction \& computations offloading}      & \checkmark & \checkmark & \checkmark& \checkmark                                                                                         \\ \hline
	\end{tabular}%
	\label{tabcompstudies}
	\vspace{-0.2cm}
\end{table*}

\subsubsection{Vehicular Fog Computing: In Brief}
Today’s modern vehicles are usually equipped with resource-rich computing and storage resources \cite{autonomousvehicleresources}. For instance, the current Tesla Model S system-on-a-chip (SoC) is equipped with 12 ARM Cortex A72 64-bit CPUs running at 2.2 GHz, and a GPU capable of 600 GFLOPS running at 1 GHz. Such a rise in computing resources gave birth to the idea of “vehicular fog computing (VFC)” \cite{vfcfirst}, enabling a cluster of parked vehicles or even closely moving vehicles as a computing infrastructure to provide cloud computing resources at the network edge. Compared to moving vehicles, parked vehicles are considered a more stable computing infrastructure since they do not change their location over a certain (long) period. Thus, by collaborating through various onboard wireless communication technologies, these parked vehicles can be seen as giant computing machinery to the outside world, providing an abundance of computing resources in the proximity of consumer vehicles. Moreover, as parked vehicles can often be found in urban areas such as shopping malls, street parking, and roadside parking lots, they can communicate with vehicular edge terminals through a dedicated gateway node, alleviating the computation burden on vehicular edge in dense traffic scenarios. The proposed FoggyEdge framework utilizes a fog of parked vehicles on top of the vehicular edge and enables efficient computation offloading to/from a vehicular fog. Additionally, as vehicular fog infrastructure may require computing services (services code) to execute the computations, the FoggyEdge framework employs the microservices concept as described in the following subsection.

\subsubsection{Microservices: In a Nutshell}
When developing automotive applications, either in-car software systems or the computing services utilized in vehicular edge/cloud computing, a conventional modular approach is usually employed, splitting the various application functionalities into small modules. Such a modularization approach is used at the time of application development and once the development phase is completed, these modules are combined under a single artefact called a monolith. In monolithic application architecture development, a module cannot be independently deployed or migrated. For instance, if a highly requested module is required to be migrated from vehicular cloud to edge, a complete instance of the entire monolithic application must be migrated, utilizing unnecessary bandwidth, and increasing the delay as well as the chances of congestion.

The research efforts to overcome these issues, especially in distributed edge cloud environments, have paved the way for the development of Microservices Architecture (MSA) \cite{microservices}\cite{microservicesbenifits}. In MSA, the modules are not only independent at the code level but they can also be independently deployed and migrated and communicated through different networking protocols. The adoption of MSA in vehicular edge cloud services and in-car software applications may result in various potential advantages such as (1) independent service development (2) continuous service integration, and (3) flexible microservice distribution to in-car electronic control units (ECUs), vehicular edge or cloud servers.

Although the MSA has the potential to decrease the complexity of existing monolithic-based vehicular services, nevertheless the underlying communication mechanism to fetch the desired microservices from the cloud utilizes inefficient address-based protocols such as wireless access in vehicular environment (WAVE) and TCP/IP \cite{connectedvehicles}. Whereas, in contrast, most vehicular services exhibit a content-centric nature due to time and space relevance (e.g., location-based services or road congestion information), and would benefit from the emerging NDN paradigm that allows applications, services, and networks to interact using the information as the main primitive. The proposed FoggyEdge framework, therefore, is built at the top of NDN principles which are described in the following subsection.

\subsubsection{Named Data Networking: An Overview}
In NDN \cite{NDN}, the communication principles are based on the application-defined content names. The consumer (i.e., the smart vehicle) forwards the request (Interest packet) to fetch the named content from the producer (i.e., vehicular edge/cloud) that generates and forwards the Data packet back to the consumer. The following are fundamental principles of NDN.

{\begin{enumerate}
    \item \textbf{ Hierarchical and Semantically Meaningful Naming Scheme:} The Interest/Data packets in NDN carries and are identified through hierarchical and semantically meaningful names.
    
    \item \textbf{Name-Based, Stateful Forwarding:} In NDN, the Interest packet heading toward the producer leaves a name-based state at each intermediate forwarder, later employed by the Data packet to reach the consumer(s). To achieve stateful forwarding, the NDN forwarders are equipped with three data structures: 1) forwarding information base (FIB) that contains multiple named entries, where a single FIB entry comprises of name prefix and corresponding interface information and is used to forward the Interest packet towards producer node, 2) pending Interest table (PIT) that stores the incoming and outgoing information of the recently forwarded Interest packets, as a result, maintain a network state to these Interests and 3) content store (CS) that caches the recently received Data packets to satisfy the similar requests in future.
    
    \item \textbf{Content-Centric Security:} To secure the content in transit across the network and at rest, each NDN Data packet carries the producer identity in the form of a digital signature.
\end{enumerate}}

\section{Related Studies}
\label{sect-related}
\begin{figure*}
		\centering
		\includegraphics[width=0.75\linewidth]{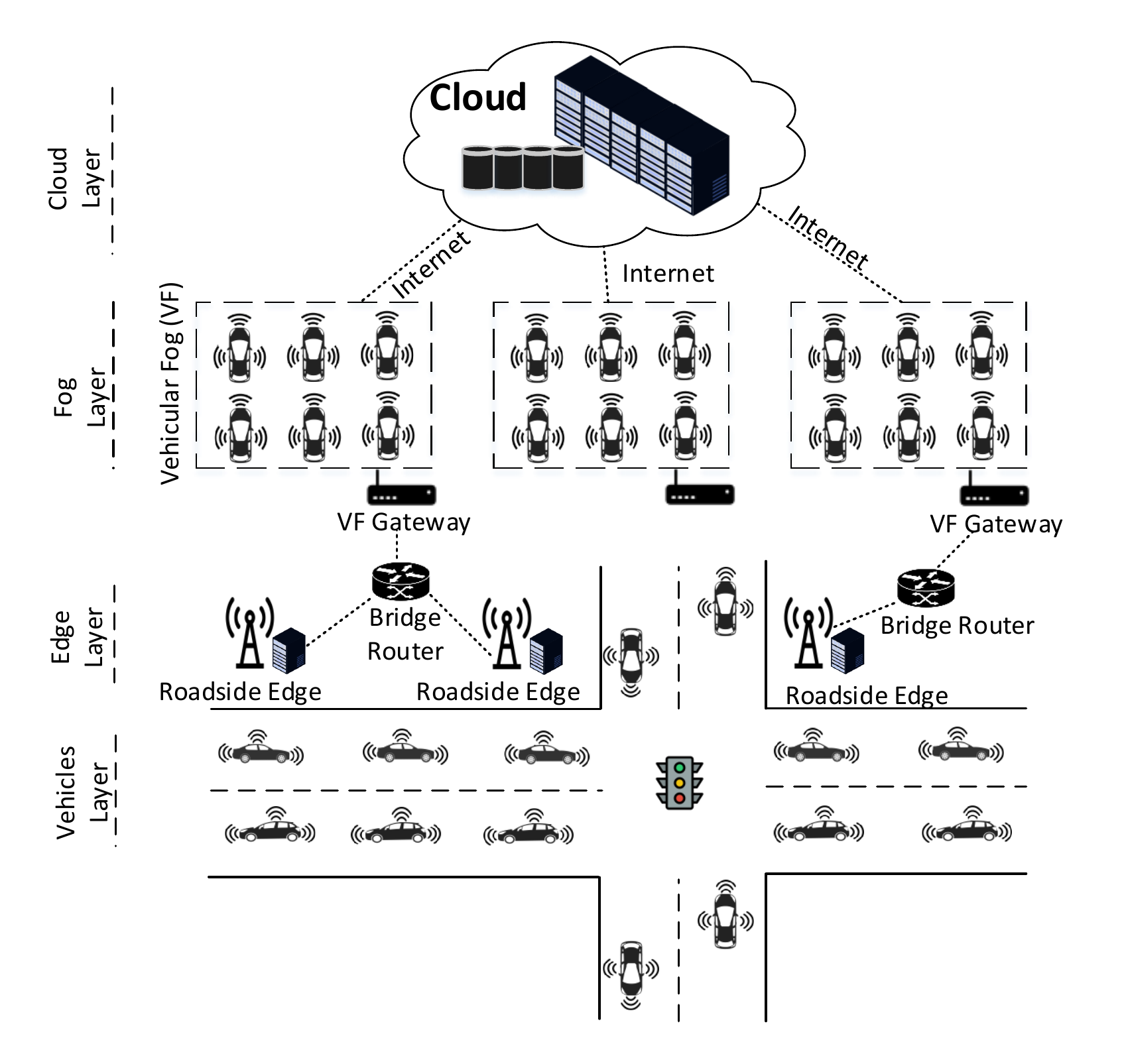}
		\caption{Proposed FoggyEdge Framework.}
		\label{fig:foggyedge}
\end{figure*}

Table \ref{tabcompstudies}  highlights the summary of related studies on vehicular fog and/or edge computing. Zhang et al, \cite{mecvehicle} proposed a cloud-based mobile edge computing offloading framework for vehicular networks aiming to effectively improve the computation transfer strategies under the considerations of task execution time and vehicle mobility. Mekki et al. \cite{MEC}  integrated multi-access edge computing with fog computing and proposed a four-layered architecture that provides computational resources in the close vicinity of end-users. A three-layered VFC framework is constructed in \cite{VFCTraffic} that aims at minimizing the latency in distributed traffic scenarios. In addition, by leveraging parked and moving vehicles, the authors also formulated VFC-enabled computation offloading scheme as an optimization problem. To improve computation efficiency in mobile vehicular clouds, a robust task scheduling mechanism considering the unstable and heterogeneous computing resources is developed in \cite{CVC}. To support latency-sensitive and computationally intensive services on the Internet of Vehicles, the authors in \cite{SDNEV} utilized software-defined networking concepts and integrated mobile edge computing with fixed edge computing to meet the latency requirements. Few other studies similar to \cite{icniov}, although built at the top of ICN communication principles and utilize vehicular fog computing concepts, significantly vary from the proposed scheme objectives and design principles i.e., the computation offloading, microservices access management mechanism, the utilization of vehicular fog at the top of the vehicular edge, and vehicle orchestration and admission process, among various other key details. 

\textbf{\textit{How does FoggyEdge Framework differ from prior works?}}
The vehicular fog/edge computing techniques outlined thus far are primarily based on the {\revision inefficient } address-centric communication architectures—the IP/WAVE—that pose significant challenges in vehicular networks owing to intermittent connectivity of mobile nodes and one-to-one stable connection requirements of IP. Moreover, the prior works do not support the Microservice access mechanism which is one of the crucial constructs in service-oriented networking. The FoggyEdge framework, in contrast, goes beyond utilizing the traditional networking protocols, for the first time utilizes the NDN as an underlying communication architecture. To this end, the FoggyEdge features the necessary modifications in the NDN protocol to support not only the named-based efficient computation offloading strategies but also provide a name-centric Microservice access management mechanism. In addition, the FoggyEdge framework, different from prior works, exploits vehicular fog at the top of the vehicular edge and thus alleviates the computation offloading selection complexity from end-user vehicles, which is a key limitation in almost all prior works.

\section{Proposed FoggyEdge Framework}
\label{sec-proposed}

So far, vehicular fog and edge computing have been mainly considered as either similar concepts or standalone solutions for computations in vehicular networks. However, they are not similar or competitive solutions, but rather complementary computing frameworks that if combined may turn out to be the game-changer for the futuristic vehicular computing framework. Therefore, going beyond the traditional solutions, in this article we propose a framework referred to as FoggyEdge, featuring a strong interplay of edge computing at the roadside units, and fog of computing at the vehicular fog layer as illustrated in Figure \ref{fig:foggyedge} and described as follows:

\textbf{ \textit{Things (Vehicles) Layer:}} The things layer consists of various consumer vehicles, equipped with limited computing and storage capacity, moving on the road, and communicating over a wireless ad hoc interface.

\textbf{ \textit{ Edge Layer:}} Edge layer consists of multiple static roadside units equipped with limited resources, referred to as vehicular edge. The key responsibility of vehicular edge is to accept computation requests from consumer vehicles over a wireless ad hoc interface and process them within the deadline time.

\textbf{ \textit{Vehicular Fog Layer:}}  This layer consists of multiple parked vehicles in the proximity of the vehicular edge, equipped with limited storage and computational capacity. The accumulation of computational and storage resources of these vehicles can be viewed as single giant computing machinery to the outside world.

\textbf{ \textit{Cloud Layer:}}  The cloud layer consists of resource-rich servers deployed at multiple hops away from the consumer vehicles moving on the road.

Before diving deep into the details of the computation offloading process, it is worth discussing the preliminary key aspects such as microservices naming schemas and access mechanisms, considering the use case of modern vehicles. The following Section sheds light on these details.

\begin{figure*}
		\centering
		\includegraphics[width=0.75\linewidth]{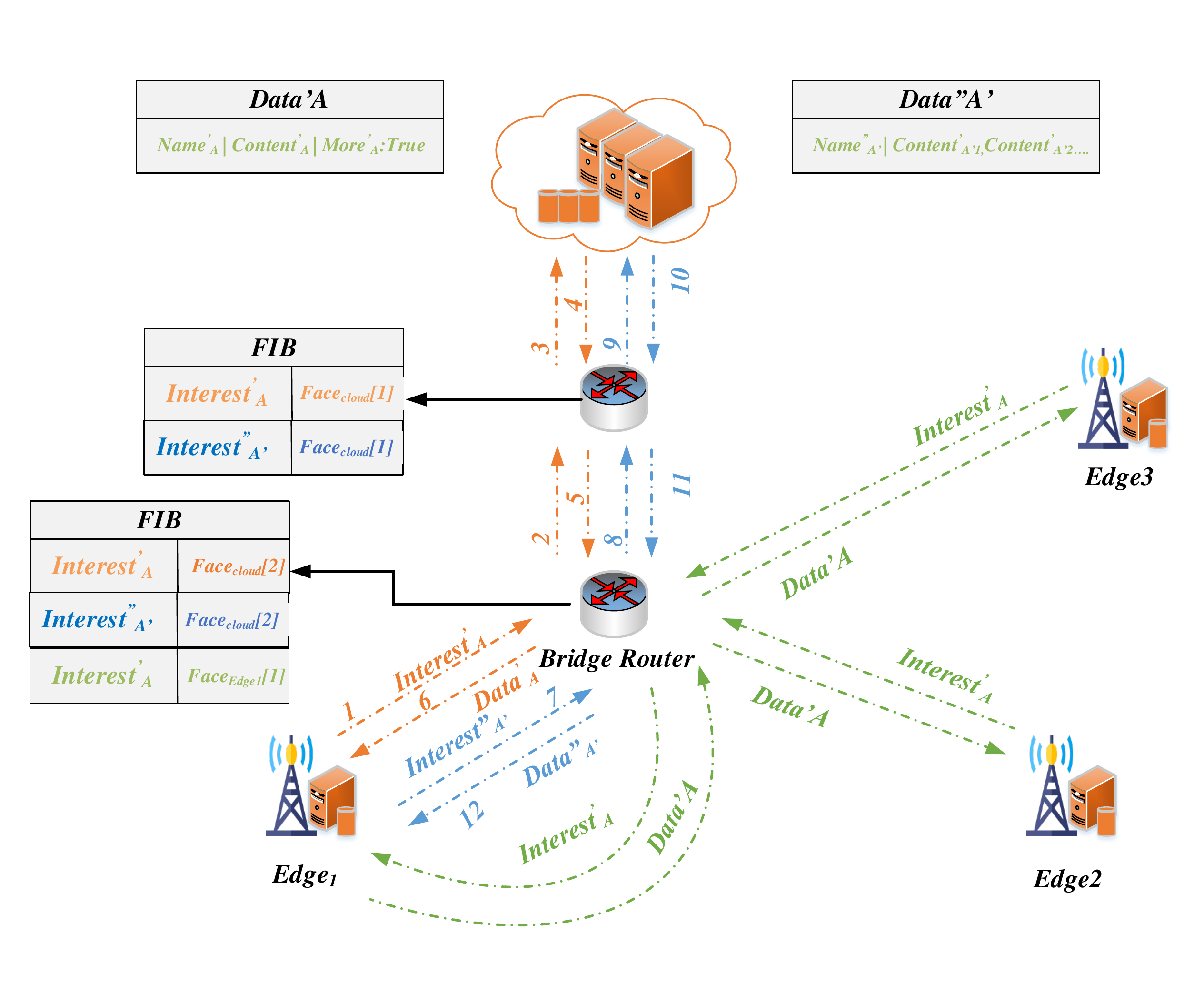}
		\caption{HMM Synchronization Process:  A Combinational Approach}
		\label{fig:hmm}
\end{figure*}

\subsection{Microservices in FoggyEdge}
The computing frameworks often require efficient management of computing services, and therefore the FoggyEdge framework employs MSA for service orchestration and management. A consumer vehicle may request various types of microservices ranging from simple microservice such as traffic condition \& monitoring microservice to more complex microservice such as a microservice for augmented reality-based image processing. When requesting a microservice, a consumer vehicle utilizes our specifically tailored naming scheme that incorporates regional information, microservice name, and the input parameters, as outlined in the following subsection.

\subsubsection{Microservices Naming Schema}

Microservices naming is the most crucial construct of the FoggyEdge framework. The way a microservice is named has a profound impact on microservices discovery, computation offloading, microservice migration, and result retrieval. In FoggyEdge, the microservices are named by following the hierarchical and semantically meaningful philosophy of NDN. To name these microservices, the proposed framework includes the regional information in the Interest packet name, in addition to the microservice name. For instance, the following name:$ FE:/Korea/Seoul/Itaewon | traffic\_status? param1, param2$, contains regional information of the country: Korea, city: Seoul, and city district: Itaewon. These three components play an important role in forwarding and/or offloading the compute requests and can be constructed by employing the built-in global positioning system (GPS) module and services similar to reverse geocoding. To this end, the city district Itaewon represents that the current compute request should be forwarded (if offloading from the vehicular edge to fog) towards a vehicular fog that is in the current vehicular edge proximity and also resides in the Itaewon area. Similarly, the city of Seoul indicates that if a vehicular fog gateway offloads the request, then it should be forwarded to the cloud node that is deployed in Seoul city. Finally, the country Korea represents the centralized cloud location for overall country-wide traffic and vehicle management and analysis. After the regional information, a vertical pipe sign (“$\vert$”) is employed to separate the regional information from the microservice name. The microservice name is defined along with the input parameters and separated by the “?” where each input parameter is further separated by a comma (“,”).

\subsubsection{Microservice Access Management Mechanism}
The proposed framework features a microservices access management mechanism, enabling the microservices to be consumed only if the consumer vehicle has legitimate rights. For instance, if a certain microservice can only be accessed after validating the access rights, the vehicular edge, in that case, may demand access rights information from the consumer vehicle. {\revision To this end, when the consumer vehicle requests the microservice using an Interest packet, access rights information is included as a secure hash-based message authentication code (HMAC). The HMAC is generated using unique components of the microservice name and smart vehicle identity with a random salt. The access rights information is added to a new field in the Interest packet named \textit{access\_rights}.}

When the vehicular edge receives an Interest packet containing the \textit{access\_rights}, it searches the Access Store (AS) for an HMAC to microservice name mapping (HMM). The structure of HMM in AS follows the key-value format as “HMAC”: “microservice-name”, where HMAC is a unique key and microservice-name is a value. As the HMAC is generated by combining the microservice name with vehicle identity (17 characters long unique vehicle identification number (VIN)) as a random salt, the presence of HMM in AS indicates that the smart vehicle has legitimate rights to consume the microservice. It is worth noting that ideally, the vehicular edge should have all the HMM information beforehand so that the verification time should not affect the latency-sensitive requests. However, if the HMM is not locally available, it has to be fetched from the cloud. To optimize the HMM fetching/synchronization process, we proposed a \textbf{combinational approach}, as illustrated in Figure \ref{fig:hmm}, and is described as follows. Following the proposed approach, when requesting the HMM, the vehicular edge adds the \textit{last\_sync\_time} field in the Interest packet, indicating when the last time the vehicular edge AS synced with the cloud AS (steps 1-3). The cloud node on receiving the Interest packet prepares the HMM response and, in the meantime extracts the \textit{last\_sync\_time}, and compares it with AS entries time to check whether there are records whose time is greater than the last sync time. If exists, the cloud node sets the \textit{more\_access\_rights} (additional field in the HMM response Data packet) bit set to “True”, and forwards it back to the requesting vehicular edge (steps 4-6). The \textit{more\_access\_rights} field indicates the presence of additional mappings at the cloud that have not been fetched yet from the requesting vehicular edge. Upon receiving the HMM response Data packet, the vehicular edge forwards another Interest packet to fetch all the remaining HMMs (steps 7-12), consequently syncing its AS with the cloud.

\subsection{FoggyEdge: Computation Philosophy \& Offloading Mechanism}
As the computational resources of the smart moving vehicles are usually exhausted owing to computationally complex processing tasks, a computational offloading process is often required to fulfil the computation request within a deadline time. We classified the offloading processes into 4 main categories as illustrated in Figure \ref{fig:offloading} (case 1-4) and are described as follows. Case 1 in Figure \ref{fig:offloading} is an obvious computation offloading from smart consumer vehicles to the vehicular edge. The vehicular edge in this case has enough computational resources and can respond within the deadline time. The remaining three cases are thoroughly discussed in the following subsection.

\subsubsection{The Offloading Process}
\begin{figure*}
		\centering
		\includegraphics[width=0.75\linewidth]{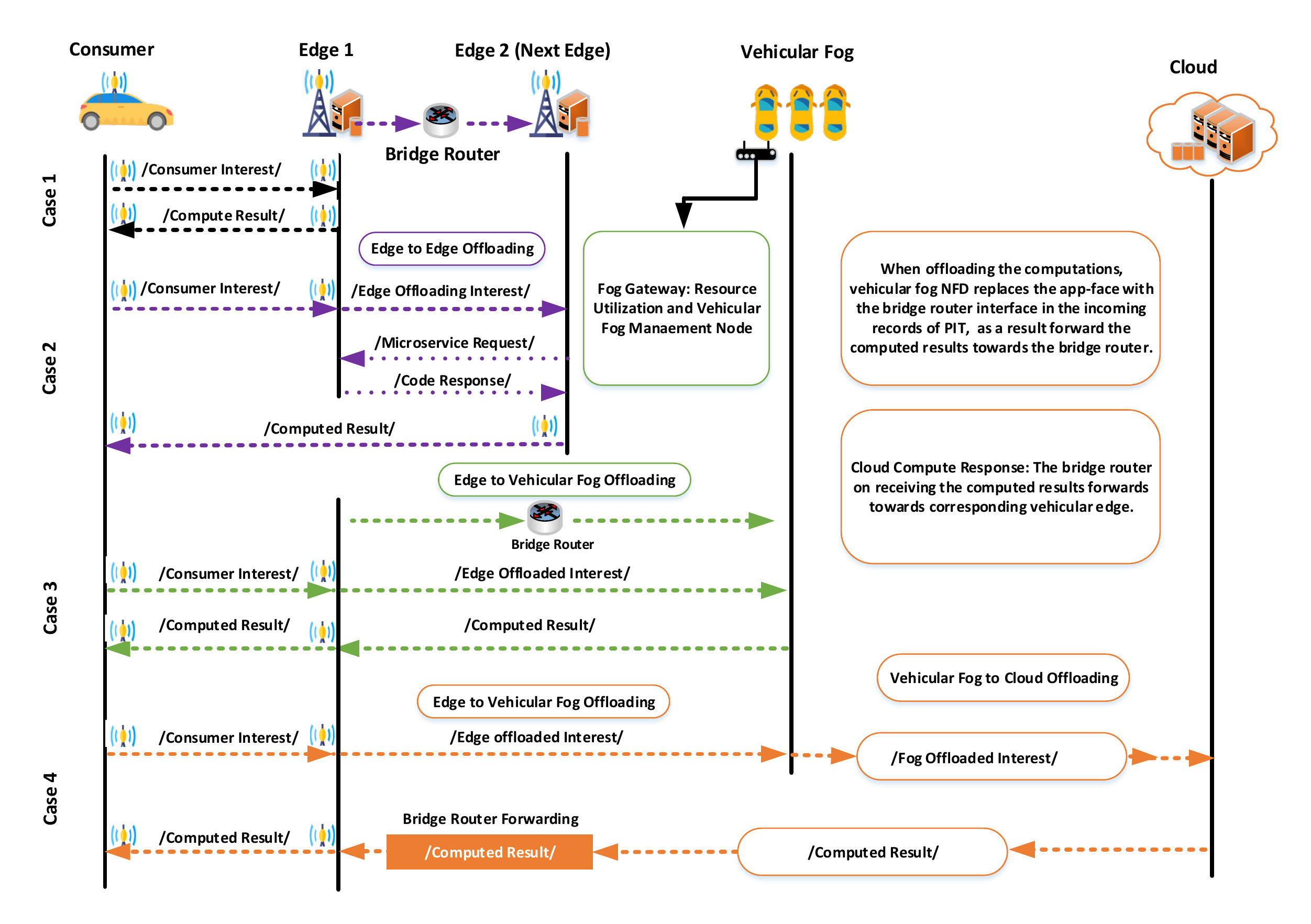}
		\caption{Compute Offloading Process }
		\label{fig:offloading}
\end{figure*}

Moving to the other cases in Figure \ref{fig:offloading}, let us assume that the vehicular edge either does not have enough computational resources or the smart consumer vehicle speed is high and soon will be out of the communication range of the vehicular edge. In both of these cases, the vehicular edge offloads the computations by adding three additional fields in the Interest packet: 1) offloading field, 2) \textit{adhoc\_response} field, and 3) \textit{microservice\_availabilty} field. When set to “True” by the vehicular edge application, the offloading field prevents the vehicular edge NDN forwarding daemon (NFD) from dropping the packet due to duplication. Thus, the Interest packet is forwarded to the next node which in our framework is the bridge router. The \textit{adhoc\_response} field indicates that the computation results must be forwarded on the wireless ad hoc interface of the “next vehicular edge” rather than the incoming interface, hence enabling the smart consumer vehicles to receive the response on their wireless ad hoc interface. The \textit{microservice\_availabilty} field when set to “True” represents that the microservice code is available on the current vehicular edge.

 The bridge router, on receiving the Interest packet that contains the aforementioned fields, employs our newly designed FoggyEdge bridge strategy for further forwarding decisions. These decisions are based on the two newly designed tables named 1) virtual edge compute FIB (VEC-FIB), and 2) virtual fog compute FIB (VFC-FIB). Among these tables, the VEC-FIB is used to keep track of all requests that are recently offloaded from the vehicular edge layer, whereas the VFC-FIB maintains a record of recently offloaded requests towards the fog layer and are yet to bring back the computation results. After deciding,  the bridge strategy either forwards the offloaded request towards the next vehicular edge (Edge to Edge offloading) or to the vehicular fog (Edge to Vehicular Fog offloading). A brief description of these offloading processes is presented below. 

\textit{EDGE TO EDGE OFFLOADING:} The bridge strategy forwards the Interest packet towards the next vehicular edge only if the VEC-FIB does not contain named entries, indicating that enough computational resources are available at the next vehicular edge. Upon receiving the offloaded Interest packet, the next vehicular edge fetches the microservice code from the previous edge---if unavailable locally, by setting the \textit{microservice\_fetch} field to “True” in the Data packet. On the way back to the previous edge, in contrast to the conventional processing, the Data packet does not purge the PIT entries but rather swaps the incoming and outgoing interfaces, hence developing the reverse PIT (R-PIT). The R-PIT mechanism keeps the PIT entries inside the intermediate nodes between the next edge and the previous edge; later consumed by the previous edge to forward the microservice code to the next edge. The following example sheds more light on the R-PIT overall process. For example, Edge 1 in Figure 3 offloads the computations towards Edge 2. Edge 2 on receiving the offloaded-computation request may request the microservice code—if unavailable locally--from Edge 1 by sending the Data packet back to Edge 1.  The Data packet acts both as an acknowledgement of offloaded Interest and a request to fetch microservice code. The Data packet processing on the way back, in contrast to the conventional processing, does not purge PIT entries; rather swaps the incoming and outgoing interfaces to create PIT entries from Edge 1 to Edge 2. In so doing, the incoming interface becomes outgoing, and the outgoing interface becomes incoming and as a result, a forwarding state is maintained that is later consumed to forward the microservice code towards Edge 2. Finally, after receiving the microservice code and performing the computations, Edge 2 forwards the computed results on its wireless ad hoc interface—if the \textit{adhoc\_response} value is “True”. Case 2 in Figure \ref{fig:offloading} illustrates the Edge-to-Edge offloading process.

\begin{figure*}
		\centering
		\includegraphics[width=0.75\linewidth]{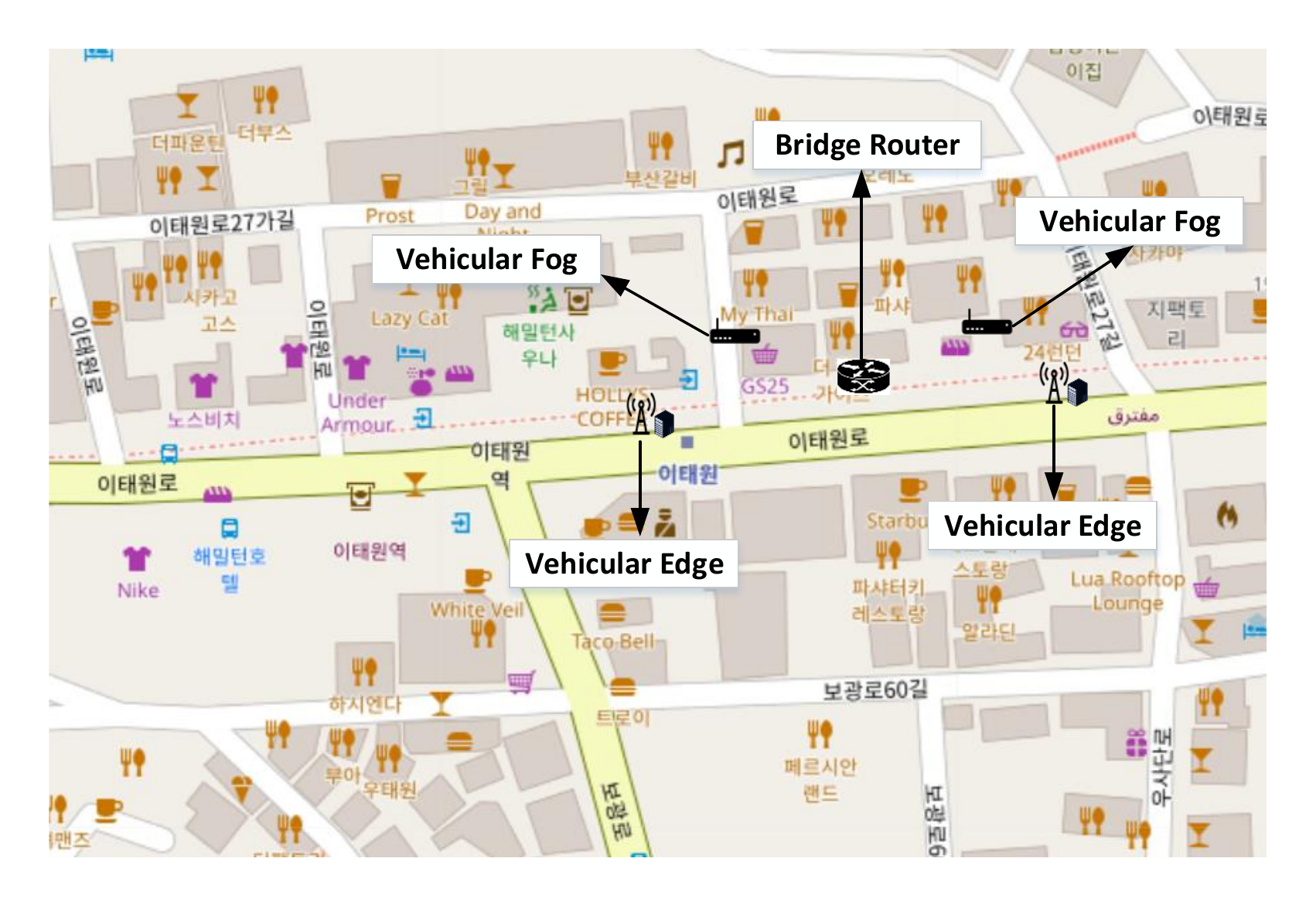}
		\caption{ Itaewon map downloaded from OpenStreetMap}
		\label{fig:openstreetmap}
\end{figure*}

\textit{EDGE TO VEHICULAR FOG OFFLOADING:}  In the presence of VEC-FIB entries for the next vehicular edge, the bridge strategy checks the VFC-FIB to select the optimal vehicular Fog. A vehicular fog performing comparatively a smaller number of computations has a higher chance to be selected. It is to note that the vehicular fog in our framework is accessible through the vehicular fog gateway (VFG). The VFG on receiving the Interest packet preserves the \textit{adhoc\_response} value, includes it in the Data packet when the results are ready, and finally forwards the response to the incoming interface, i.e., towards the bridge router. The bridge router, upon receiving the Data packet extracts the \textit{adhoc\_response} field value; if set to “True” forwards the Data packet toward the next vehicular edge. The next-edge on receiving the Data packet forwards it to the wireless ad-hoc interface which is then received by the consumer vehicle. Case 3 in Figure \ref{fig:offloading} illustrates the Edge-to-Fog offloading process.

\textit{VEHICULAR FOG TO CLOUD OFFLOADING:} Vehicular fog to cloud offloading enables the VFG to offload the computations to the cloud node in the case when adequate vehicular computational resources are not available in VFG. Case 4 in Figure \ref{fig:offloading} illustrates the Fog-to-Cloud offloading process. Similar to the vehicular edge, the VFG sets the offloading field to “True” to enable the VFG NFD to forward the packet toward the cloud. When the computed result arrives from the cloud, the VFG forwards the packet towards the bridge router, which then follows the same process as mentioned in the edge-to-vehicular fog offloading scenario.

\subsection{Vehicular Fog Management}
As vehicular fog is a cluster of multiple vehicles providing computational and storage resources, the efficient vehicle orchestration mechanism is of utmost importance. To manage the resources of these vehicles, VFG maintains the vehicular fog resource access table (VF-RAT), which includes information such as available resources, the status of the current instance running on the vehicular fog node, the location of the vehicle inside the parking area, and the estimated parking time of the vehicle.

In the proposed scheme, we assumed that the VFG can receive incoming traffic from wired and wireless ad-hoc interfaces. The wired interfaces are used to connect VFG with vehicular edge and cloud nodes, whereas the wireless Adhoc interface is used for communication with parked vehicles.

\subsubsection{Vehicle Orchestration and Admission Process}
Whenever a vehicle enters the parking area and passes through the ticket zone, it sends an Interest packet on the wireless ad hoc interface, requesting the parking slot information from the VFG. While sending the Interest packet, the vehicle adds the following information: 1) estimated parking time, and 2) the available resources. Upon receiving the Interest packet, the VFG extracts the aforementioned information and selects the parking slot for the new vehicle. Soon after selecting the slot number, the VFG performs two operations: 1) sends the Data packet back to the vehicle informing the slot location, and 2) saves the vehicle information in the VF-RAT table. In so doing, the VFG maintains overall control of all vehicles and forwards the future compute request towards an optimal vehicle. In case, if any vehicle running the computations wishes to leave the parking area, offload the computations to the VFG. The offloading process is as follows. The departing vehicle generates an Interest packet on its wireless ad hoc interface, which is received by the VFG at its ad hoc interface. VFG, on receiving the offloading request, selects an optimal vehicle, inserts its identity in the Data packet, and forwards the response on its ad hoc interface. As the response is received by all vehicles including departing vehicles and newly selected vehicles, the departing vehicle forwards the partially computed instance towards the new vehicle and finally leaves the parking area. As soon as the departing vehicle passes through the ticket zone, the VFG erases all of its information from the VF-RAT.

\begin{figure*}
		\centering
		\includegraphics[width=0.7\linewidth]{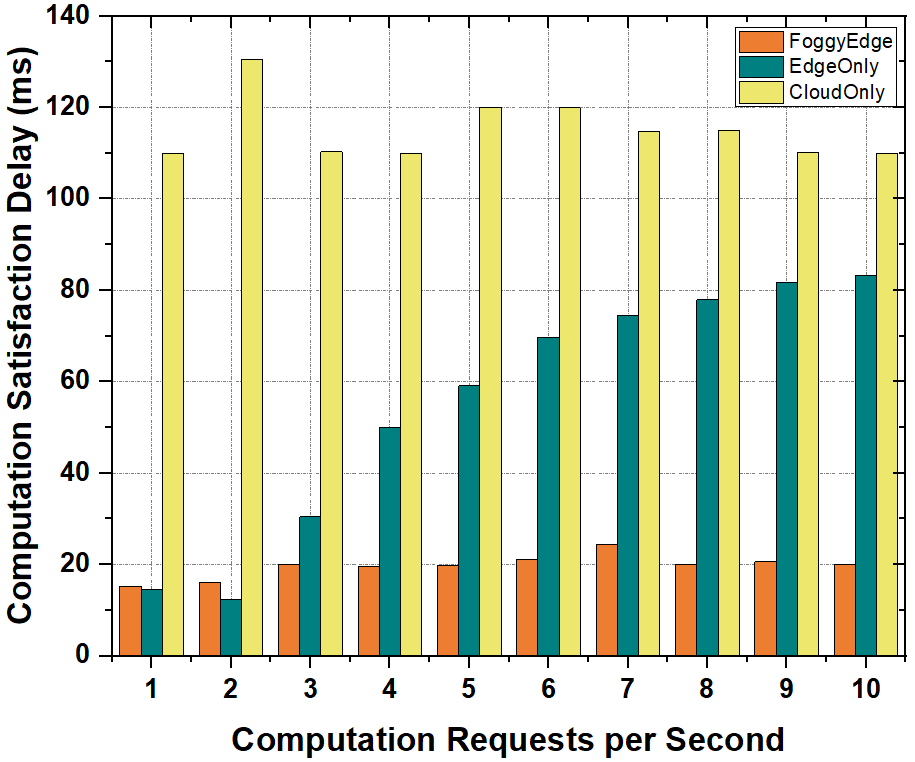}
		\caption{ CSD as a function of number of computation requests per seconds}
		\label{fig:result}
\end{figure*}
\section{Preliminary Performance Evaluation}
\label{sec-performance}
\subsection{Simulation Setup}
To validate the effectiveness of the FoggyEdge framework, a preliminary comparative performance analysis based on simulation of urban mobility (SUMO) \cite{sumo} and OpenStreetMap (OSM) \cite{open-street} is conducted in the ndnSIM simulator \cite{ndnSIM}. We deploy vehicular edge and vehicular fog nodes (VFG is accumulating the fog resources) on the Itaewon road of Seoul city in South Korea which we exported from OSM as illustrated in Figure 4. It is worth mentioning here that the Itaewon region is only selected to impersonate a realistic road infrastructure scenario; our framework, instead, is applicable to any urban vehicular scenario. Moreover, to mimic the computational behaviour of Microservices at different layers, we exploit our open-sourced ndnCSIM codebase \footnote{source code available at: https://github.com/11th-ndn-hackathon/ndn-compute-simulator } that implements the basic requirements of microservices execution and efficient management of node’s resources. {\revision In terms of node's resources, the file "n3/network/model/node.h" in the ndnSIM simulator is updated to echo a realistic scenario of edge/fog/cloud nodes resources. To this end, new properties are added, including $m\_initial\_compute\_resources$ which represents a node's initial computation resources in units; $m\_compute\_resources$ which denotes the remaining computation resources after microservices occupy compute resources for execution; and $m\_initial\_compute\_capacity$ depicting a node's remaining computational capacity as a percentage. The updated Node Class i.e., the "n3/network/model/node.h" file enables resource allocation to edge/fog/cloud nodes in "units". During simulation, we assign values such as 1000 units to $m\_initial\_compute_resources$ corresponding to real-world computation resources of a node, for instance, 4 GHz. Moreover, at the start of the simulation, both $m\_compute\_resources$ and $m\_initial\_compute\_resources$ are initialized with the same values. However, as the microservice request arrives, $m\_compute\_resources$ is reduced in accordance with the computation necessities of the requested microservice. Finally, once the microservice release resource, the $m\_compute\_resources$ value is updated and the available computational resource is increased.}

In addition to the aforementioned changes, we also modify the NFD codebase to incorporate the necessary requirements of computation offloading. For the evaluation, the microservices-based computation requests are generated at different rates ranging from 1 request/second to 10 requests/second, mimicking the excessive load behaviour on the vehicular edge. A total of five different types of microservices, in terms of computational requirements, are employed in the simulation and among them, each microservice resource requirement varies from the others. The microservices selection in each request follows the random distribution. Furthermore, to ensure a fair comparison, we selected NDN-based EdgeOnly and CloudOnly schemes rather than IP-based address-centric proposals outlined in the related studies Section. The rationale is that the address-centric schemes lack key network performance enrichment features such as 1) request aggregation, 2) in-network caching, and 3) multi-casting support, consequently intrinsically performing worst compared to name-centric proposals \cite{connectedvehicles}.

\subsection{Computation Satisfaction Delay:} 
We measure the computations satisfaction delay (CSD) of the FoggyEdge framework compared to the EdgeOnly and CloudOnly solutions as shown in Figure \ref{fig:result}. The CSD specifies the time it takes 1) for a computation request to reach the computing node (i.e., vehicular edge, vehicular fog, or cloud server), 2) for the computing node to process the request based on the microservice type, and 3) the corresponding computation results to reach the requesting consumer vehicle. 

In comparison, our results demonstrate that the CSD, when employing the FoggyEdge framework, is lower than the other two solutions in most cases. Analyzing the results, it can be seen that when the requests generation rate is low (i.e., at 1 request/second and 2 requests/seconds), the CSD for FoggyEdge and EdgeOnly is almost the same. The reason for such performance is that when the generation rate is low, almost all the requests are satisfied directly through the vehicular edge, without being offloaded to further layers. On the contrary, when the request generation rate is increasing and the vehicular edge is offloading the requests, a significant difference in CSD can be observed for these three solutions. Delivering proximate computations due to the inclusion of a vehicular fog layer, stabilize the computations offloading from the edge, and is the main reason behind low CSD in the FoggyEdge framework compared to EdgeOnly and CloudOnly solutions.

\section{Future Research Directions}
\label{sec-future}
The integration of VFC with VEC by considering microservices for computations and service migration, and the NDN as an underlying communication architecture is still in its infancy and requires further research efforts in this direction, as briefly outlined in the following subsections.

\subsection{Incentive Mechanism in FoggyEdge} 
Naturally, vehicle owners may refuse to provide the computational resources due to battery power consumption concerns. Similarly, if the parking lot owners are deploying the gateway nodes, they may expect to receive a satisfying reward in return for their services. Therefore, designing an incentive mechanism in FoggyEdge, that benefits both the parked vehicle owners and the parking lot owners, is essential to effectively meet the computational demands of end users’ smart vehicles. These incentives can collaboratively be provided by the government authorities and the vendors of smart vehicles requesting the computations. To this end, a blockchain-based incentive mechanism can be designed that not only comply with the incentive requirements but also operates on top of the NDN protocol. 

\subsection{Security and Privacy} 

The computation requests offloaded by the end user’s smart vehicles are usually privacy-sensitive and therefore enabling privacy-preserving mechanisms as well as secure communication is important for the complete realization of the FoggyEdge framework. Different from traditional IP-based security mechanisms, an NDN-based secure communication and privacy-preserving framework are required when offloading the computation requests from 1) smart vehicles to vehicular edge, 2) from vehicular edge to vehicular fog, 3) and finally from vehicular fog to cloud nodes.  

\subsection{Optimal Vehicular Fog Location} 
A drastic increase in smart vehicles especially in urban areas is escalating the number of parked vehicles which can be found in abundance in the proximity of vehicular edges such as in shopping malls, offices, and restaurants. A question may arise about whether it is cost-effective to deploy vehicular fog infrastructure within all these parking places. Though the deployment cost for vehicular fog will be an order of magnitude less compared to the vehicular edge deployment, nevertheless, a handful amount of financial resources will be required, making the optimal location of vehicular fog infrastructure deployment an important challenge to address.

\subsection{Load Balancing and Optimal Vehicle Selection For Computations} 

To preserve the energy consumption and to fairly utilize the computational resources of fog vehicles, an efficient load balancing scheme considering the computation resources requirements of the incoming requests and the available resources of fog vehicles is of utmost importance. To this end, a resource-aware optimization scheme can be designed that fairly divides the task among available vehicles and consequently improves the reliability and effectiveness of vehicular fog computations.

\section{Conclusion}
\label{sec-conclusion}
To provide computations in the proximity of consumer vehicles, this article presents an information-centric vehicular fog-enabled computation offloading and management scheme. Specifically, a four-layer FoggyEdge framework is proposed to enable dynamic cooperation among multiple layers for computations load balancing. The benefits of the proposed framework are as follows:
\begin{enumerate}
    \item	Following the NDN principles, FoggyEdge naturally benefits from the features such as request aggregation, in-network caching, and multicasting.
    
    \item The integration of vehicular fog at layer 3 essentially alleviates the computation burden from the vehicular  edge and cloud.
    
    \item 3)	The response time can be significantly reduced as the vehicular fog is in proximity to the vehicular edge.

\end{enumerate}
To the best of our knowledge, the proposed framework is the first to integrate vehicular fog nodes at the top of the vehicular edge and follows the NDN principles.  Finally, as a part of future work, we intend to perform a comprehensive performance analysis of FoggyEdge considering various other key metrics including computation satisfaction rate, microservices code migration time using R-PIT, and computation offloading overhead within the vehicular fog.

\section*{Acknowledgment}

This research was financially supported in part by the National Research Foundation of Korea(NRF) grant funded by the Korea government(MSIT) (No. 2022R1A2C1003549). Moreover, it is also partially supported by the US National Science Foundation (award CNS-2306685).

\ifCLASSOPTIONcaptionsoff
  \newpage
\fi

\bibliographystyle{IEEEtran}
\bibliography{foggyedge}

\begin{IEEEbiography}[{\includegraphics[width=1in,height=1.25in,clip,keepaspectratio]{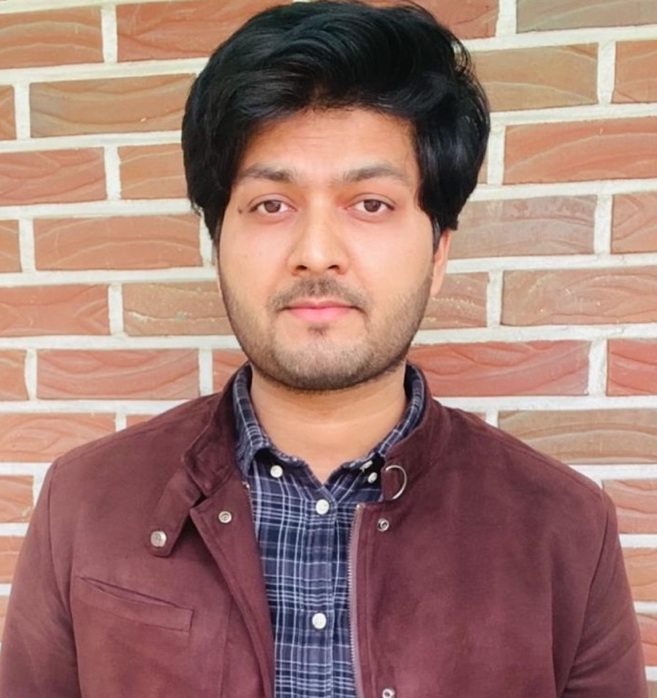}}]{Muhammad Atif Ur Rehman } is a Lecturer (Assistant Professor) in the Department of Computing and Mathematics at the Manchester Metropolitan University, UK. He received his Ph.D. in Electronics and Computer Engineering from Hongik University, South Korea in 2022. His research interests include distributed networking and computing, metaverse-based networking and computing, edge/fog computing, and intelligent network protocol designing. 
\end{IEEEbiography}

\begin{IEEEbiography}[{\includegraphics[width=1in,height=1.25in,clip,keepaspectratio]{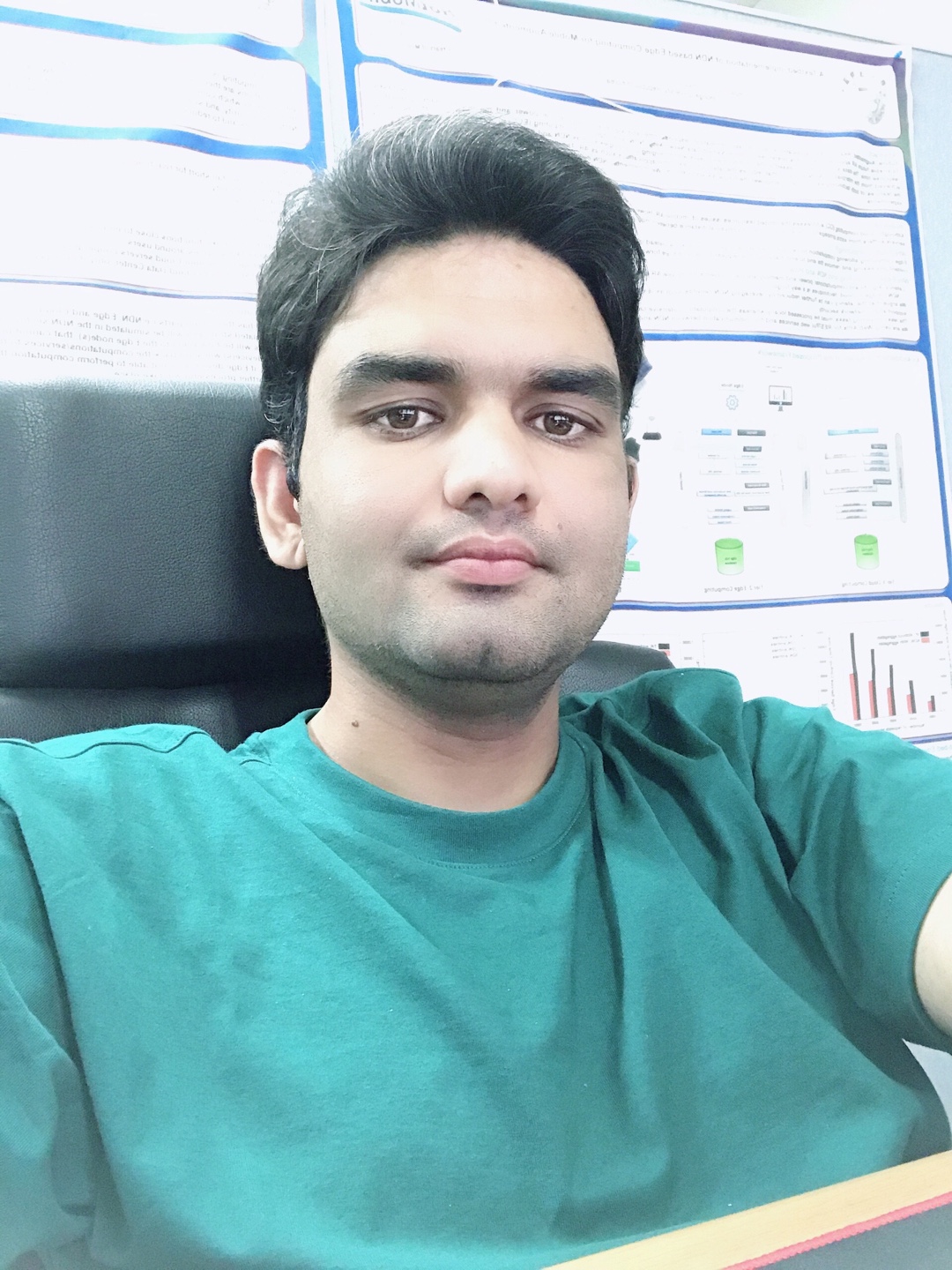}}]{Muhammad Salahuddin  (Student Member, IEEE)} received M.S. degree in Computer Science from COMSATS Islamabad University in 2016. He is currently pursuing Ph.D. in Computer Engineering student from Hongik University, South Korea. His research interests include WSNs, Vehicular networks,  ICN/NDN, Intelligent Edge/Fog computing, metaverse, and IoTs.
\end{IEEEbiography}

\begin{IEEEbiography}[{\includegraphics[width=1in,height=1.25in,clip,keepaspectratio]{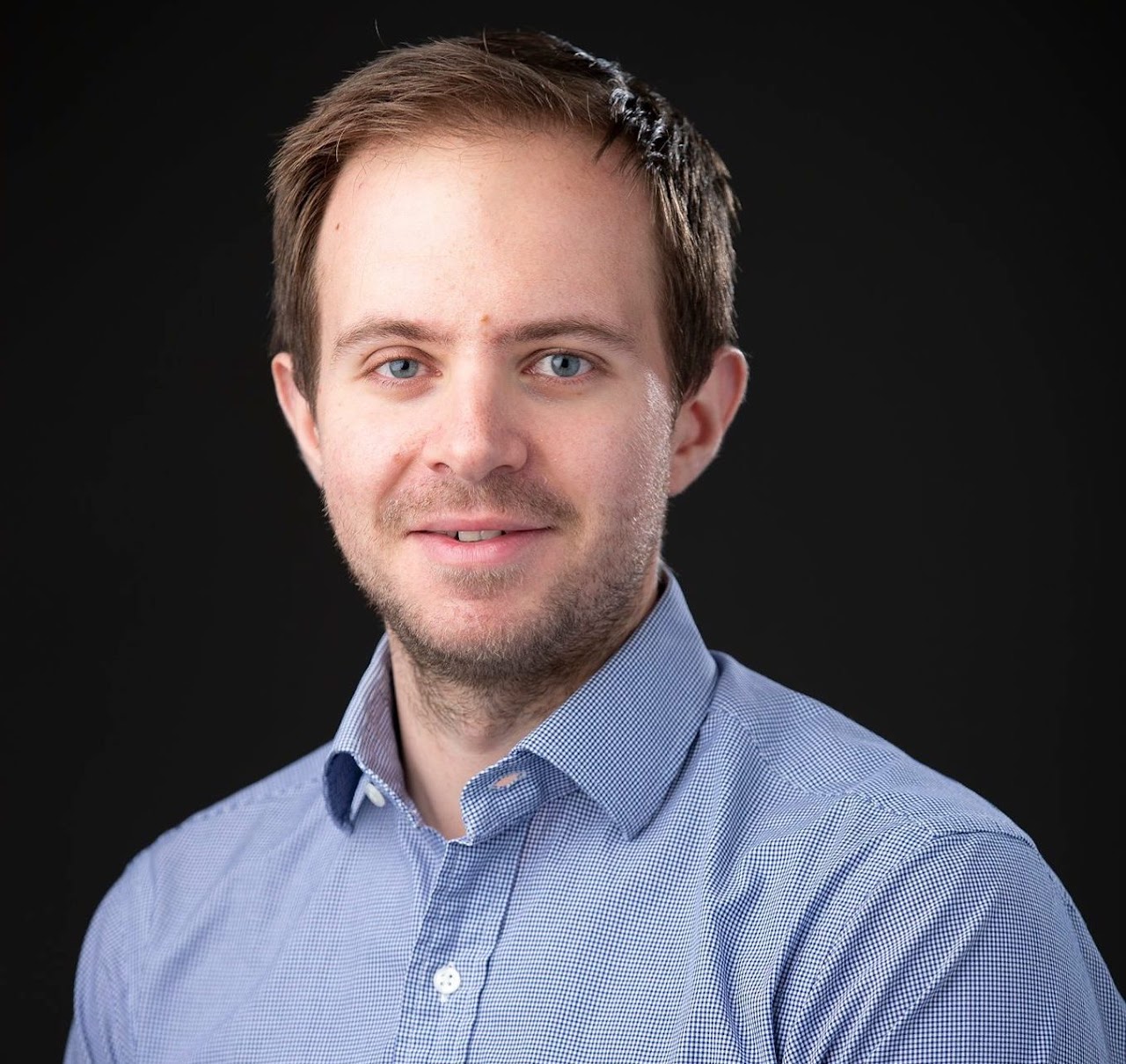}}]{Spyridon Mastorakis  (mastorakis@nd.edu, Member, IEEE)}  is an Assistant Professor in Computer Science and Engineering at the University of Notre Dame. He received his Ph.D. in Computer Science from the University of California, Los Angeles in 2019. His research interests include network systems and architecture, edge computing, and security.

\end{IEEEbiography}

\begin{IEEEbiography}[{\includegraphics[width=1in,height=1.25in,clip,keepaspectratio]{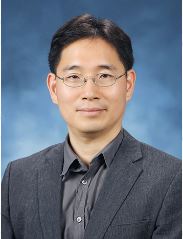}}]{Byung-Seo Kim (M’02-SM’17)}

is a Full Professor in the Software and Communications Engineering Department, at Hongik University, South Korea. He received his PhD in electrical and computer engineering from the University of Florida, Gainesville, FL, USA, in 2004. His research interests include designing and developing efficient wireless/wired networks, including wireless CCNs/NDNs, and mobile-edge computing. 
\end{IEEEbiography}

\end{document}